\ifdefined\llncs
\documentclass[envcountsame]{llncs}
\else
\documentclass[reqno]{amsart}
\fi

\input macros.tex

\begin{document}

\title{The equational theory of \\ the natural join and inner union
  is decidable} 
\ifdefined\llncs
\author{Luigi Santocanale
  \\\email{luigi.santocanale@lif.univ-mrs.fr}} 
\institute{LIF, CNRS UMR 7279, Aix-Marseille Universit\'e}
\else
\author[Luigi Santocanale]{Luigi Santocanale}
\address{Luigi Santocanale\\
LIF, CNRS UMR 7279, Aix-Marseille Universit\'e}
\email{luigi.santocanale@lif.univ-mrs.fr}
\fi

\maketitle

\begin{abstract}
The natural join and the inner union operations combine relations of a
database. Tropashko and Spight \cite{Tropashko2008} realized that
these two operations are the meet and join operations in a class of
lattices, known by now as the relational lattices. They proposed then
lattice theory as an algebraic approach to the theory of databases,
alternative to the relational algebra.

Previous works \cite{LitakMHjlamp,RAMICS17} proved that the
quasiequational theory of these lattices---that is, the set of
definite Horn sentences valid in all the relational lattices---is
undecidable, even when the signature is restricted to the pure lattice
signature.

We prove here that the equational theory of relational lattices is
decidable. That, is we provide an algorithm to decide if two lattice
theoretic terms $t,s$ are made equal under all interpretations in some
relational lattice. We achieve this goal by showing that if an
inclusion $t \leq s$ fails in any of these lattices, then it fails in
a relational lattice whose size is bound by a triple exponential
function of the sizes of $t$ and $s$.

\end{abstract}

\section{Introduction}

The natural join and the inner union operations combine relations
(i.e. tables) of a database. SQL-like languages construct queries by
making repeated use of the natural join and of the union. The inner
union is a mathematically well behaved variant of the union---for
example, it does not introduce empty cells.  Tropashko and Spight
realized \cite{Tropashko2005,Tropashko2008} that these two operations
are the meet and join operations in a class of lattices, known by now
as the class of relational lattices. They proposed then lattice theory
as an algebraic approach, alternative to Codd's relational algebra
\cite{codd70}, to the theory of databases.

Roughly speaking, elements of the relational lattice $\R(D,A)$ are
tables of a database, where $A$ is a set of columns' names and $D$ is
the set of possible cells' values.
Let us illustrate the two operations with examples. The \nj
takes two tables and constructs a new one whose columns are indexed by
the union of the headers, and whose rows are glueings of the rows
along identical values in common columns:
\begin{center}
  \table{ll}{
    Author & Area \\ \hline
    Santocanale & Logic  \\
    Santocanale & CS   
  }
  \;$\bowtie$\;
  \table{ll}{
    Area & Reviewer \\\hline
    CS & Turing  \\
    Logic & G\"odel}
  \;=\; \table{lll}{
    Author& Area & Reviewer \\\hline
    Santocanale & Logic & G\"odel \\
    Santocanale & CS & Turing }
\end{center}
The \iu restricts two tables to the common columns and lists all the
rows of the two tables. The following example suggests how to
construct, using this operation, a table of users given two (or more)
tables of people having different roles.
\begin{center}
  \table{lll}{
    \multicolumn{3}{|c|}{Author}\\
    \hline Name & Surname & Conf \\ \hline
    Luigi & Santocanale & FOSSACS }
  \quad$\cup$\quad
  \table{lll}{
    \multicolumn{3}{|c|}{Reviewer}\\ \hline 
    Name & Surname & Area \\\hline
    Alan & Turing & CS \\
    Kurt & G\"odel & Logic }
  \quad=\quad \table{ll}{
    \multicolumn{2}{|c|}{User}\\ \hline 
    Name & Surname \\\hline
    Luigi & Santocanale  \\
    Alan & Turing \\
    Kurt & G\"odel }
\end{center}
Since we shall focus on lattice-theoretic considerations, we shall use
the symbols $\land$ and $\vee$, in place of the symbols $\bowtie$ for
$\cup$ used by database theorists.

A first important attempt to axiomatize these lattices was done by
Litak, Mikul\'as, and Hidders \cite{LitakMHjlamp}. They proposed an
axiomatization, comprising equations and quasiequations, in a
signature that extends the pure lattice signature with a constant, the
header constant. A main result of that paper is that the
quasiequational theory of relational lattices is undecidable in this
extended signature. Their proof mimics Maddux's proof that the
equational theory of cylindric algebras of dimension $n \geq 3$ is
undecidable \cite{Maddux1980}.

Their result was further refined by us in \cite{RAMICS17}: the
quasiequational theory of relational lattices is undecidable even when
the signature considered is the least one, comprising only the meet
(natural join) and the join operations (inner union). Our proof relied
on a deeper algebraic insight: we proved that it is undecidable
whether a finite \si lattice can be embedded into a relational
lattice---from this kind of result, undecidability of the
quasiequational theory immediately follows. We proved the above
statement by reducing to it an undecidable problem in modal logic, the
coverability problem of a frame by a universal $\Sfive^{3}$-product
frame \cite{HHK2002}. In turn, this problem was shown to be
undecidable by reducing it to the representability problem of finite
simple relation algebras \cite{HH2001}.

We prove here that the equational theory of relational lattices is
decidable. That is, we prove that it is decidable whether two lattice
terms $t$ and $s$ are such that $\eval[v]{t} = \eval[v]{s}$, for any
valuation $v : \X \rto \RDA$ of variables in a relational lattice
$\R(D,A)$.  We achieve this goal by showing that this theory has a
kind of finite model property of bounded size. Out main result,
Theorem~\ref{theo:main1}, sounds as follows: \emph{if an inclusion
  $t \leq s$ fails in a relational lattice $\R(D,A)$, then such
  inclusion fails in a finite lattice $\R(E,B)$, such that $B$ is
  bound by an exponential function in the size of $t$ and $s$, and $E$
  is linear in the size of $t$.} It follows that the size of $\R(E,B)$
can be bound by a triple exponential function in the size of $t$ and
$s$.  In algebraic terms, our finite model theorem can be stated by
saying that the variety generated by the relational lattices is
actually generated by its finite generators, the relational lattices
that are finite.

In our opinion, our results are significant under two main
respects. Firstly, the algebra of the natural join and of the inner
union is so spread and used via the SQL-like languages.  We dare to
say that most of programmers that use a database---more or less
explicitly, for example within server-side web programs---are using
these operations. In view of the widespread use of this algebraic
system, its decidability status deserved being settled. Moreover, we
believe that the mathematical insights contained in our decidability
proof shall contribute to understand further the algebraic system. For
example, it is not known yet whether a complete finite axiomatic basis
exists for relational lattices; finding it could eventually yield
applications, e.g. on the side of automated optimization of queries.

Secondly, our work exhibits the equational theory of relational
lattices as a decidable one within a long list of undecidable logical
theories \cite{Maddux1980,HH2001,HHK2002,LitakMHjlamp,RAMICS17} that
are used used to model the constructions of relational algebra.  We
are exploring limits of decidability, a research direction widely
explored in automata theoretic settings starting from \cite{caucal}.
We do this, within logic and with plenty of potential applications,
coming from the undecidable side and crossing the border: after the
quasiequational theory, undecidable, the next natural theory on the
list, the equational theory of relational lattices, is decidable.

On the technical side, our work relies on \cite{San2016} where the
duality theory for finite lattices developed in \cite{San09:duality}
was used to investigate equational axiomatizations of relational
lattices. A key insight from \cite{San2016} is that relational
lattices are, in some sense, duals of generalized ultrametric spaces
over a powerset algebra. It is this perspective that made it possible
to uncover the strong similarity between the lattice-theoretic methods
and tools from modal logic---in particular the theory of combination
of modal logics, see e.g. \cite{Kurucz2007}. We exploit here this
similarity to adapt filtrations techniques from modal logic
\cite{Gabbay} to lattice theory. Also, the notion of generalized
ultrametric spaces over a powerset algebra and the characterization of
injective objects in the category of these spaces have been
fundamental tools to prove the undecidability of the quasiequational
theory \cite{RAMICS17} as well as, in the present case, the
decidability of the equational theory.

\smallskip

The paper is organised as follows. We recall in
Section~\ref{sec:defselconcepts} some definitions and facts about
lattices. The relational lattices $\RDA$ are introduced in
Section~\ref{sec:rellattices}.
In Section~\ref{sec:spaces} we show how to construct a lattice
$\L(X,\d)$ from a generalized ultrametric space $(X,\d)$. This
contruction generalizes the construction of the lattice $\RDA$: if
$X = \AD$ is the set of all functions from $A$ to $D$ and $\d$ is as a
sort of Hamming distance, then $\L(X,\d) = \RDA$.  We use the
functorial properties of $\L$ to argue that when a finite space
$(X,\d)$ has the property of being \Pc, then $\L(X,\d)$ belongs to the
variety generated by the relational lattices.
In Section~\ref{sec:tableaux} we show that if an inclusion $t \leq s$
fails in a lattice $\RDA$, 
then we can construct a finite subset $T(f,t) \subseteq \AD$, a
``tableau" witnessing the failure, such that if $T(f,t) \subseteq T$
and $T$ is finite, then $t \leq s$ fails in a finite lattice of the
form $\L(T,\d_{B})$, where the distance $\d_{B}$ takes values in a
finite powerset algebra $P(B)$.
In Section~\ref{sec:completion}, we show how to extend $T(f,t)$ to a
finite bigger set $\GT$, so that $(\GT,\d_{B})$ as a space over the
powerset algebra $P(B)$ is \Pc. This lattice $\L(\GT,\d_{B})$ fails
the inclusion $t \leq s$; out of it, we build a lattice of the form
$\R(E,B)$, which fails the same inclusion; the sizes of $E$ and $B$
can be bound by functions of the sizes of the terms $t$ and $s$.
Perspectives for future research directions appear in the last
Section~\ref{sec:conclusions}.

\section{Elementary notions on orders and lattices}
\label{sec:defselconcepts}

We assume some basic knowledge of order and lattice theory as
presented in standard monographs \cite{DP02,GLT2}. Most of the lattice
theoretic tools we use originate from the monograph \cite{FJN}.

A \emph{lattice} is a poset $L$ such that every finite non-empty
subset $X \subseteq L$ admits a smallest upper bound $\bv X$ and a
greatest lower bound $\bigwedge X$.  A lattice can also be understood
as a structure $\mathfrak{A}$ for the functional signature
$(\vee,\land)$, such that the interpretations of these two binary
function symbols both give $\mathfrak{A}$ the structure of an
idempotent commutative semigroup, the two semigroup structures being
connected by the absorption laws $x \land (y \vee x) = x$ and
$x \vee (y \land x) = x$.  Once a lattice is presented as such
structure, the order is recovered by stating that $x \leq y$ holds if
and only if $x \land y= x$.

A lattice $L$ is \emph{complete} if any subset $X \subseteq L$ admits
a smallest upper bound $\bv X$. It can be shown that this condition
implies that any subset $X \subseteq L$ admits a greatest lower bound
$\bigwedge X$. A lattice is \emph{bounded} if it has a least element
$\bot$ and a greatest element $\top$. A complete lattice (in
particular, a finite lattice) is bounded, since $\bigvee \emptyset$
and $\bigwedge \emptyset$ are, respectively, the least and greatest
elements of the lattice.

If $P$ and $Q$ are partially ordered sets, then a function
$f : P \rto Q$ is \emph{\op} (or \emph{monotone}) if $p \leq p'$
implies $f(p) \leq f(p')$.  If $L$ and $M$ are lattices, then a
function $f : L \rto M$ is a \emph{lattice morphism} if it preserves
the lattice operations $\vee$ and $\land$. A lattice morphism is
always \op. A lattice morphism $f : L \rto M$ between bounded lattices
$L$ and $M$ is \emph{\bp} if $f(\bot) = \bot$ and $f(\top) = \top$.  A
function $f : P \rto Q$ is said to be \emph{\la} to an \op
$g : Q \rto P$ if $f(p) \leq q$ holds if and only if $p \leq g(q)$
holds, for every $p \in P$ and $q \in Q$; such a \la, when it exists,
is unique. Dually, a function $g : Q \rto P$ is said to be \emph{\ra}
to an \op $f : P \rto Q$ if $f(p) \leq q$ holds if and only if
$p \leq g(q)$ holds; clearly, $f$ is \la to $g$ if and only if $g$ is
\ra to $f$, so we say that $f$ and $g$ form an adjoint pair. If $P$
and $Q$ are complete lattices, the property of being a \la (resp.,
\ra) to some $g$ (resp., to some $f$) is equivalent to preserving all
(possibly infinite) joins (resp., all meets).

A \emph{Moore family on $P(U)$} is a collection $\F$ of subsets
of $U$ which is closed under arbitrary intersections.  Given a Moore
family $\F$ on $P(U)$, the correspondence sending
$Z \subseteq U$ to
$\closure{Z}:= \bigcap \set{Y \in \F \mid Z \subseteq Y }$ is a
\emph{closure operator} on $P(U)$, that is, an \op inflationary and
idempotent endofunction of $P(U)$.  The subsets in $\F$, called
the \emph{closed sets}, are exactly the fixpoints of this closure
operator.  A Moore family $\F$ has the structure of a \cl where
\begin{align}
  \label{eq:opsMooreFamily}
  \bigwedge
  \, X & := \bigcap X\,, &
  \bigvee
  \,X & := \closure{\bigcup X}\,.
\end{align}
The notion of Moore family can also be defined for an arbitrary \cl
$L$.  Moore families on $L$ turns out to be in bijection with closure
operators on $L$. We shall actually consider the dual notion: a
\emph{dual Moore family on a complete lattice $L$} is a subset
$\F\subseteq L$ that is closed under arbitrary joins. Such an
$\F$ determines an interior operator (an \op decreasing and
idempotent endofunction on $L$) by the formula
$\interior{x} = \bigvee \set{y \in \F \mid y \leq x}$ and has
the structure of a complete lattice, where
$\bigvee_{\F} X := \bigvee_{L} X$ and
$\bigwedge_{\F} X := \interior{(\bigwedge_{L} X)}$.
Dual Moore families on $L$ are in bijection with interior operators on
$L$.  Finally, let us mention that closure (resp., interior)
operators, whence Moore families (resp., dual Moore families), arise
from adjoint pairs $f$ and $g$ (with $f$ \la to $g$) by the formula
$\closure{x} = g(f(x))$ (resp., $\interior{x} = f(g(x))$).

\section{The relational lattices $\R(D,A)$}
\label{sec:rellattices}

Throughout this paper we use the  $\expo{X}{Y}$ for the
set of functions of domain $Y$ and codomain $X$.

Let $A$ be a collection of attributes (or column names) and let $D$ be
a set of cell values. A \emph{relation} on $A$ and $D$ is a pair
$(\alpha,T)$ where $\alpha \subseteq A$ and
$T \subseteq \expo{\alpha}{D}$.  Elements of the relational lattice
$\R(D,A)$
\footnote{In \cite{LitakMHjlamp} such a lattice is called \emph{full}
  relational lattice. The wording ``class of relational lattices'' is
  used there for the class of lattices that have an embedding into
  some lattice of the form $\R(D,A)$.  }
are relations on $A$ and $D$.
Informally, a relation $(\alpha,T)$ represents a table of a relational
database, with $\alpha$ being the header, i.e. the collection of names
of columns, while $T$ is the collection of rows.

Before we define the \nj, the \iu operations, and the order on
$\R(D,A)$, let us recall some key operations. If
$\alpha \subseteq \beta \subseteq A$ and $f \in \expo{\beta}{D}$, then
we shall use $f \restr[\alpha] \in \expo{\alpha}{D}$ for the
restriction of $f$ to $\alpha$; if $T \subseteq \expo{\beta}{D}$, then
$T \rrestr[\alpha]$ shall denote projection to $\alpha$, that is, the
direct image of $T$ along restriction,
$T \rrestr[\alpha] := \set{ f \restr[\alpha] \mid f \in T}$; if
$T \subseteq \expo{\alpha}{D}$, then $i_{\beta}(T)$ shall denote
cylindrification to $\beta$, that is, the inverse image of
restriction,
$i_{\beta}(T) := \set{ f \in \expo{\beta}{D} \mid f_{\restriction
    \alpha} \in T}$.  Recall that $i_{\beta}$ is \ra to
$\rrestr[\alpha]$.  With this in mind, the \nj and the inner union of
relations are respectively described by the following formulas:
\begin{align*}
  (\alpha_{1},T_{1}) \land (\alpha_{2},T_{2})
  & := (\alpha_{1} \cup \alpha_{2},T)  \\
  \text{where }T & = \set{f \mid f \restr[\alpha_{i}] \in T_{i}, i =
    1,2} \\
  & = i_{\alpha_{1} \cup \alpha_{2}}(T_{1}) \cap i_{\alpha_{1} \cup
    \alpha_{2}}(T_{2})\,,  \\
  (\alpha_{1},T_{1}) \vee (\alpha_{2},T_{2})
  & := (\alpha_{1} \cap \alpha_{2},T) \\
  \text{where }T & = \set{f \mid \exists i\in \set{1,2},\exists
    g \in T_{i} \tst g\, \restr[\alpha_{1} \cap \alpha_{2}] = f} \\
  & = T_{1} \rrestr[\alpha_{1} \cap \alpha_{2}] \cup \,T_{2}
  \rrestr[\alpha_{1} \cap \alpha_{2}]\,.
\end{align*}
The order is then given by 
\choosedisplay{$(\alpha_{1},T_{1}) \leq (\alpha_{2},T_{2}) \tiff \alpha_{2} \subseteq \alpha_{1}
  \tand T_{1} \rrestr[\alpha_{2}] \subseteq T_{2}$.}{
\begin{align*}
  (\alpha_{1},T_{1}) & \leq (\alpha_{2},T_{2}) \quad \tiff \quad \alpha_{2} \subseteq \alpha_{1}
  \tand T_{1} \rrestr[\alpha_{2}] \subseteq T_{2}\,.
\end{align*}
}

A convenient way of describing these lattices was introduced in
\cite[Lemma 2.1]{LitakMHjlamp}. The authors argued that the relational
lattices $\R(D,A)$ are isomorphic to the lattices of closed subsets of
$A \cup \AD$, where $Z \subseteq A \cup \AD$ is said to be closed if
it is a fixed-point of the closure operator $\closure{(\,-\,)}$
defined as
\begin{align*}
  \closure{Z} & := Z \cup \set{f \in \AD \mid A \setminus Z \subseteq
    Eq(f,g), \text{ for some $g \in Z$} }\,,
\end{align*}
where in the formula above $Eq(f,g)$ is the equalizer of $f$ and
$g$. Letting
\choosedisplay{$\d(f,g) := \set{x \in A \mid f(x) \neq g(x)}$,}{
\begin{align*}
  \d(f,g) & := \set{x \in A \mid f(x) \neq g(x)}\,,
\end{align*}
}
the above definition of the closure operator is obviously equivalent
to the following one:
\begin{align*}
  \closure{Z} & := \alpha \cup \set{f \in \AD \mid \d(f,g) \subseteq
    \alpha, \text{ for some $g \in Z \cap \AD$} },\;
  \text{with $\alpha = Z \cap A$}.
\end{align*}
From now on, we rely on this representation of relational lattices.

\section{Lattices from metric spaces}
\label{sec:spaces}

Generalized ultrametric spaces over a Boolean algebra $P(A)$ turn out
to be a convenient tool for studying relational lattices
\cite{LitakMHjlamp,San2016}. 
Metrics are well known
tools from graph theory, see e.g. \cite{HIK2012}.  Generalized
ultrametric spaces over a Boolean algebra $P(A)$ were introduced in
\cite{PriessCrampeRibenboim1995} to study equivalence relations.

\begin{definition}
  An \emph{ultrametric space over $P(A)$} (briefly, a \emph{space}) is a pair
  $(X,\d)$, with $\d : X \times X \rto P(A)$ such that, for every
  $f,g,h \in X$,
  \begin{align*}
    \delta(f,f) & \subseteq \emptyset\,, &
    \delta(f,g) & \subseteq \d(f,h) \cup \d(h,g)\,.
  \end{align*}
\end{definition}
We have defined an ultrametric space over $P(A)$ as a category (with a
small set of objects) enriched over $(P(A)^{op},\emptyset,\cup)$, see
\cite{lawvere}. We assume in this paper that such a space $(X,\d)$ is
also \emph{reduced} and \emph{symmetric}, that is, that the following
two properties hold for every $f,g \in X$:
\begin{align*}
  \d(f,g) & = \emptyset \text{ implies } f = g, & 
  \d(f,g) & =
  \d(g,f)\,.
\end{align*}

A \emph{morphism} of spaces\footnote{As $P(A)$ is not totally ordered,
  we avoid calling a morphism ``\emph{non expanding map}'' as it is
  often done in the literature.}  $\psi : (X,\d_{X}) \rto (Y,\d_{Y})$
is a function $\psi: X \rto Y$ such that
$\d_{Y}(\psi(f),\psi(g)) \leq \d_{X}(f,g)$, for each $f,g \in X$.
Obviously, spaces and their morphisms form a category.  If
$\d_{Y}(\psi(f),\psi(g)) = \d_{X}(f,g)$, for each $f,g \in X$, then
$\psi$ is said to be an \emph{isometry}.
A
space $(X,\d)$ is said to be \emph{\Pc}, see \cite{Ackerman2013}, or
\emph{convex}, see \cite{Pouzet}, if, for each $f,g \in X$ and
$\alpha,\beta \subseteq A$,
\begin{align*}
  \delta(f,g) \subseteq \alpha \cup \beta& \text{ implies }
  \delta(f,h) \subseteq \alpha \text{ and } \delta(h,g) \subseteq
  \beta\,,
  \;\text{ for some $h \in X$}.
\end{align*}
\begin{proposition}[see \cite{PriessCrampeRibenboim1995,Ackerman2013}]
  If $A$ is finite, then a space is injective in the category of
  spaces if and only if it is \Pc.
\end{proposition}

If $(X,\d_{X})$ is a space and
$Y \subseteq X$, then the restriction of $\d_{X}$ to $Y$ induces a
space $(Y,\d_{X})$; we say then that $(Y,\d_{X})$ is a \emph{subspace}
of $X$.  Notice that the inclusion of $Y$ into $X$ yields an isometry
of spaces.

Our main example of space over $P(A)$ is $(\AD,\d)$, with $\AD$ the
set of functions from $A$ to $D$ and the distance defined by
\begin{align}
  \label{def:distance}
  \d(f,g) & := \set{ a \in A \mid f(a) \neq g(a)}\,.
\end{align}
A second example is a slight generalization of the previous one.
Given a surjective function $\pi : D \rto A$, let $\Secpi$ denote the
set of all the functions $f : A \rto D$ such that
$\pi \circ f = id_{A}$. Then $\Secpi \subseteq \AD$, so $\Secpi$ with
the distance inherited from $(\AD,\d)$ can be made into a space.
Considering the first projection $\pi_{1} : A \times D \rto A$, we see
that $(D^{A},\d)$ is isomorphic to the space $\Sec{\pi_{1}}$.
By identifying $f \in \Secpi$ with a vector
$\langle f(a) \in \pi^{-1}(a)\mid a \in A\rangle$, we see that
\begin{align}
  \label{eq:sepuniversalproductframe}
  \Secpi & = \prod_{a \in A}D_{a}\,, \quad\text{where
    $D_{a} := \pi^{-1}(a)$. }
\end{align}
That is, the spaces of the form $\Secpi$ are naturally related to
Hamming graphs in combinatorics \cite{MR1788124}, dependent function
types in type theory \cite{jacobs,dyckhoff}, universal
$\Sfive^{A}$-product frames in modal logic \cite{HHK2002}.
\begin{theorem}[see \cite{RAMICS17}]
  Spaces of the form $\Secpi$ are, up to isomorphism, exactly the
  injective objects in the category of spaces.
\end{theorem}

\subsection{The lattice of a space}

The construction of the lattice $\RDA$ can be carried out from any
space.  Namely, for a space $(X,\d)$ over $P(A)$, say that
$Z \subseteq X$ is $\alpha$-closed if $g \in Z$ and
$\d(f,g) \subseteq \alpha$ implies $f \in Z$.
Clearly, $\alpha$-closed subsets of $X$ form a Moore family so, for
$Z \subseteq X$, we denote by $\closure[\alpha]{Z}$ the least closed
subset of $X$ containing $Z$. Observe that $f \in \closure[\alpha]{Z}$
if and only if $\d(f,g) \subseteq \alpha$ for some $g \in Z$.
Next and in the rest of the paper, we shall exploit the obvious
isomorphism between $P(A)\times P(X)$ and $P(A \cup X)$ (where we
suppose $A$ and $X$ disjoint) and notationally identify a pair
$(\alpha,Z) \in P(A)\times P(X)$ with its image
$\alpha\cup X \in P(A \cup X)$. Let us say then that $(\alpha,Z)$ is
closed if $Z$ is $\alpha$-closed.  Closed subsets of $P(A \cup X)$
form a Moore family, whence a \cl where the order is subset inclusion.
\begin{definition}
  For a space $(X,\d)$, the lattice $\L(X,\d)$ is the lattice of
  closed subsets of $P(A \cup X)$.
\end{definition}
Clearly, for the space $(\AD,\d)$, we have $\L(\AD,\d) = \RDA$. Let us
mention that meets and joins $\L(X,\d)$ are computed using the
formulas in~\eqref{eq:opsMooreFamily}. In particular, for joins,
\begin{align*}
  (\alpha,Y) \vee (\beta,Z) & = (\alpha \cup \beta, \closure[\alpha
  \cup \beta]{Y \cup Z})\,.
\end{align*}
The above formula yields that, for some $f \in X$,
$f \in (\alpha,Y) \vee (\beta,Z)$ if and only if
$\d(f,g) \subseteq \alpha \cup \beta$, for some $g \in Y \cup Z$.

We argue next that the above construction is functorial. Below, for a
a function $\psi : X \rto Y$, $\psi^{-1} : P(Y) \rto P(X)$ is the
inverse image of $\psi$, defined by
$\psi^{-1}(Z) := \set{x \in X \mid \psi(x) \in Z}$.
\begin{proposition}
  \label{prop:Lfunctor}
  If $\psi :(X,\d_{X}) \rto (Y,\d_{Y})$ is a space morphism and
  $(\alpha,Z) \in \L(Y,\d_{Y})$, then
  $(\alpha,\psi^{-1}(Z)) \in \L(X,\d_{X})$. Therefore, by defining
  $\L(\psi)(\alpha,Z) := (\alpha,\psi^{-1}(Z))$, the construction $\L$
  lifts to a contravariant functor from the category of spaces to the
  category of \cmsl{s}.
\end{proposition}
\begin{proof}
  Let $f \in X$ be such that, for some $g \in \psi^{-1}(Z)$
  (i.e. $\psi(g) \in Z$), we have $\d_{X}(f,g) \subseteq \alpha$. Then
  $\d_{Y}(\psi(f),\psi(g)) \subseteq \d_{X}(f,g) \subseteq \alpha$, so
  $\psi(f) \in Z$, since $Z$ is $\alpha$-closed, and
  $f \in \psi^{-1}(Z)$.
  In order to see that $\L(\psi)$ preserves arbitrary intersections,
  recall that $\psi^{-1}$ does.
  \qed
\end{proof}
Notice that $\L(\psi)$ might not preserve arbitrary joins. 
\begin{proposition}
  \label{prop:HSSecpi}
  The lattices $\L(\Secpi)$ generate the same lattice variety of the
  lattices $\RDA$.
\end{proposition}
That is, a lattice equation holds in all the lattices $\L(\Secpi)$ if
and only if it holds in all the relation lattices $\RDA$.
\begin{proof}
  Clearly, each lattice $\RDA$ is of the form $\L(\Secpi)$. Thus we
  only need to argue that every lattice of the form $\L(\Secpi)$
  belongs to the lattice variety generated by the $\RDA$, that is, the
  least class of lattices containing the lattices $\RDA$ and closed
  under products, sublattices, and homomorphic images. We argue as
  follows.
    
  As every space $\Secpi$ embeds into a space $(\AD,\d)$ and a space
  $\Secpi$ is injective, we have maps $\iota : \Secpi \rto \ADd$ and
  $\psi : \ADd \rto \Secpi$ such that
  $\psi \circ \iota = id_{\Secpi}$. By functoriality,
  $\L(\iota) \circ \L(\psi) = id_{\L(\Secpi)}$. Since $\L(\iota)$
  preserves all meets, it has a left adjoint
  $\ell : \L(\Secpi) \rto \L(\AD,\d) = \RDA$. It is easy to see that
  $(\ell,\L(\psi))$ is an EA-duet in the sense of \cite[Definition
  9.1]{permutohedra} and therefore $\L(\Secpi)$ is a homomorphic image of a
  sublattice of $\RDA$, by \cite[Lemma~9.7]{permutohedra}.
  \qed
\end{proof}

\begin{remark}
  For the statement of \cite[Lemma~9.7]{permutohedra} to hold,
  additional conditions are necessary on the domain and the codomain
  of an EA-duet. Yet the implication that derives being a homomorphic
  image of a sublattice from the existence of an EA-duet is still
  valid under the hypothesis that the two arrows of the EA-duet
  preserve one all joins and, the other, all meets.
\end{remark}

\subsection{Change of Boolean-algebra}

We have not distinguished yet among spaces over some $P(A)$ and spaces
over some $P(B)$. For our goals, we need to consider the following
case. We suppose that $P(B)$ is a \sBA of $P(A)$ via an inclusion
$i : P(B) \rto P(A)$. If $(X,\d_{B})$ is a space over $P(B)$, then we
can transform it into a space $(X,\d_{A})$ over $P(A)$ by setting
$\d_{A}(f,g) = i(\d_{B}(f,g))$. We have therefore two lattices
$\L(X,\d_{B})$ and $\L(X,\d_{A})$.

\begin{proposition}
  \label{prop:changeOfBA}
  Let $\beta \subseteq B$ and $Y \subseteq X$. Then $Y$ is
  $\beta$-closed if and only if it is $i(\beta)$-closed. Consequently
  the map $\is$, sending $(\beta,Y) \in \L(X,\d_{B})$ to
  $\is(\beta,Y) := (i(\beta),Y) \in \L(X,\d_{A})$, is a lattice
  embedding.
\end{proposition}
\begin{proof}
  Observe that $\d_{B}(f,g) \subseteq \beta$ if and only if
  $\d_{A}(f,g) = i(\d_{B}(f,g)) \subseteq i(\beta)$.  This immediately
  implies the first statement of the Lemma, but also that, for
  $Y \subseteq X$, $\closure[\beta]{Y} = \closure[i(\beta)]{Y}$.
  Using the fact that meets are computed as intersections and that $i$
  preserves intersections, it is easily seen that $\is$ preserves
  meets. For joins let us compute as follows:
  \begin{align*}
    \is(\beta_{1},Y_{1}) \vee \is(\beta_{2},Y_{2})
    & = (i(\beta_{1})\cup i(\beta_{2}),\closure[i(\beta_{1}) \cup i(\beta_{2})]{Y_{1} \cup Y_{2}}) \\
    = \;&  (i(\beta_{1}\cup \beta_{2}),\closure[i(\beta_{1} \cup
    \beta_{2})]{Y_{1} \cup Y_{2}})
    = (i(\beta_{1}\cup \beta_{2}),\closure[\beta_{1} \cup
    \beta_{2}]{Y_{1} \cup Y_{2}}) \\
    = \;&  \is(\beta_{1}\cup \beta_{2},\closure[\beta_{1} \cup
    \beta_{2}]{Y_{1} \cup Y_{2}})
    = \is((\beta_{1},Y_{1}) \vee (\beta_{2},Y_{2})).
    \tag*{\qed}
  \end{align*}
\end{proof}

\section{Failures from big to small lattices}
\label{sec:tableaux}

The set of lattice terms is generated by the following grammar:
\begin{align*}
  t & := x \mid \top \mid t \land t \mid \bot \mid t \vee t\,,
\end{align*}
where $x$ belongs to a set of variables $\X$. For lattice terms
$t_{1},\ldots ,t_{n}$, we use $Vars(t_{1},\ldots ,t_{n})$ to denote
the set of variables (which is finite) occurring in any of these
terms.
The size of a term $t$ is the number of nodes in the representation of
$t$ as a tree.
If $v : \X\rto L$ is a valuation of variables into a lattice $L$, the
value of a term $t$ w.r.t. the valuation $v$ is defined by induction
in the obvious way; here we shall use $\eval[v]{t}$ for it.

For $t,s$ two lattice terms, the inclusion $t \leq s$ is the equation
$t \vee s = s$.
Any lattice-theoretic equation is equivalent to a pair of inclusions,
so the problem of deciding the equational theory of a class of
lattices reduces to the problem of decing inclusions.
An inclusion $t \leq s$ is valid in a class of lattices $\Kl$ if, for
any valuation $v : \X \rto L$ with $L \in \Kl$,
$\eval[v]{v} \leq \eval[v]{s}$; it fails in $\Kl$ if for some
$L \in \Kl$ and $v : \X \rto L$ we have
$\eval[v]{t} \not\leq \eval[v]{s}$.

From now on, our goal shall be proving that if an inclusion $t \leq s$
fails in a lattice $\RDA$, then it fails in a lattice $\LSp$, where
$\Secpi$ is a finite space over some finite Boolean algebra
$P(B)$. The size of $B$ and of the space $\Secpi$, shall be inferred
from of the sizes of $t$ and $s$.

From now on, we us fix terms $t$ and $s$, a lattice $\RDA$, and a
valuation $v : \X \rto \RDA$ such that
$\eval[v]{t} \not\subseteq \eval[v]{s}$.
\begin{lemma}
  If, for some $a \in A$, $a \in \eval[v]{t} \setminus \eval[v]{s}$,
  then the inclusion $t \leq s$ fails in the lattice $\R(E,B)$ with
  $B = \emptyset$ and $E$ a singleton.
\end{lemma}
\begin{proof}
  The map sending $(\alpha,X) \in \RDA$ to $\alpha \in P(A)$ is
  lattice morphism. Therefore if $t \leq s$ fails because of
  $a \in A$, then it already fails in the Boolean lattice
  $P(A)$. Since $P(A)$ is distributive, $t \leq s$ fails in the two
  elements lattice. Now, when $B = \emptyset$ and $E$ is a singleton
  $\R(E,B)$ is (isomorphic to) the 2 elements lattice, so the same
  equation fails in $\R(E,B)$.
  \qed
\end{proof}
Because of the Lemma, we shall focus on functions $f \in \AD$ such
that $f \in \eval[v]{t} \setminus \eval[v]{s}$. In this case we shall
say that \emph{$f$ witnesses the failure of $t \leq s$} (in $\RDA$,
w.r.t. the valuation $v$).

\subsection{The lattices $\RDAT$}

Let $T$ be a subset of $\AD$ and consider the subspace $(T,\d)$ of
$\AD$ induced by the inclusion $i_{T} : T \subseteq \AD$. According to
Proposition~\ref{prop:Lfunctor}, the inclusion $i_{T}$ induces a
\cmslh $\L(i_{T}) : \RDA = \L(\AD,\d) \rto \L(T,\d)$. Such a map has a
\ra $j_{T} : \L(T,\d) \rto \L(\AD,\d)$, which is a \cjslh; moreover
$j_{T}$ is injective, since $\L(i_{T})$ is surjective.
\begin{proposition}
  \label{prop:joinsemilattice}
  For a subset $T \subseteq \AD$ and $(\alpha,X) \in \RDA$,
  $(\alpha, \closure[\alpha]{X \cap T}) =
  j_{T}(\L(i_{T}(\alpha,X))$. The set of elements of the form
  $(\alpha,\closure[\alpha]{X \cap T})$, for $\alpha \subseteq A$ and
  $X \subseteq \AD$, is a complete sub-join-semilattice of $\RDA$.
\end{proposition}
\begin{proof}
  It is easily seen that $\L(i_{T})(\alpha,X) = (\alpha, X \cap T)$
  and that, for $(\beta,Y) \in \L(T,\d)$,
  $(\beta,Y) \subseteq (\alpha,X \cap T)$ if and only if
  $(\beta,\closure[\beta]{Y}) \subseteq (\alpha,X)$, so
  $j_{T}(\beta,Y) = (\beta,\closure[\beta]{Y})$.

  It follows that the elements of the form
  $(\alpha,\closure[\alpha]{X \cap T})$, where $(\alpha,X) \in \RDA$,
  form a sub-\cjsl of $\RDA$: indeed, they are the image of lattice
  $\L(T,\d)$ under the \cjslh $j_{T}$. We argue next that, for any
  pair $(\alpha,X)$ (we do not require that $X$ is $\alpha$-closed)
  there is a $Z \subseteq \AD$ which is $\alpha$-closed and such that
  $\closure[\alpha]{X \cap T} = \closure[\alpha]{Z \cap T}$. Indeed,
  the equality
  \begin{align*}
    \clos[\alpha]{X \cap T} & =  \clos[\alpha]{\clos[\alpha]{ X \cap T} \cap T}
  \end{align*}
  is easily verified, so we can let $Z = \clos[\alpha]{X \cap T}$.
  \qed
\end{proof}

Therefore, the set of pairs of the form
$(\alpha,\closure[\alpha]{X \cap T})$ is a dual Moore family and a
complete lattice, where joins are computed as in $\RDA$, and where
meets are computed in a way that we shall make explicit. For the
moment, let us fix the notation.
\begin{definition}
  $\RDAT$ is the lattice of elements of the form form
  $(\alpha,\closure[\alpha]{X \cap T})$.
\end{definition}
By the proof of Proposition~\ref{prop:joinsemilattice}, the lattice
$\RDAT$ is isomorphic to the latttice $\L(T,\d)$.
We shall use the symbol $\bigmmeet$ for meets in $\RDAT$; these are
computed by the formula
\begin{align*}
  \bigmmeet_{i \in I} (\alpha_{i},X_{i}) & = \interior{(\bigcap_{i \in
      I} \alpha_{i}, \bigcap_{i \in I} X_{i})}\,, 
\end{align*}
where, for each $(\alpha,X) \in \RDA$, $\interior{(\alpha,X)}$ is the
greatest pair in $\RDAT$ that is below $(\alpha,X)$. Standard theory
on adjoints yields
\begin{align*}
  \interior{(\alpha,X)} = (j_{T} \circ \L(i_{T}))(\alpha,X) =
  (\alpha,\closure[\alpha]{X \cap T})\,.
\end{align*}
We obtain in this way the explicit formula for the binary meet in
$\RDAT$:
\begin{align*}
  (\alpha, \closure[\alpha]{X \cap T})
  \mmeet
  (\beta, \closure[\beta]{Y \cap T})  
  & = (\alpha \cap \beta, \closure[\alpha \cap
  \beta]{\closure[\alpha]{X \cap T}  \cap \closure[\beta]{Y \cap T}
    \cap T})\,.
\end{align*}
Remark that we have
\begin{align*}
  (\alpha,X) \mmeet (\beta,Y) & \subseteq (\alpha,X) \cap (\beta, Y)
\end{align*}
whenever $(\alpha,X)$ and $(\beta,Y)$ are in $\RDAT$.

\begin{lemma}
  \label{lemma:conj}
  Let $(\alpha,X),(\beta,Y) \in \RDAT$ and let
  $f \in T$. If $f \in (\alpha,X) \cap (\beta,Y)$, then
  $f \in (\alpha,X) \mmeet (\beta,Y)$.
\end{lemma}
\begin{proof}
  This is immediate from the fact that
  \begin{align*}
    \closure[\alpha]{X \cap T} \cap \closure[\beta]{Y \cap T} \cap T &
    \subseteq \closure[\alpha \cap \beta]{\closure[\alpha]{X \cap T}
      \cap \closure[\beta]{Y \cap T} \cap T}\,.  \tag*{\qedhere}
  \end{align*}
\end{proof}

\subsection{Preservation of the failure in the lattices $\RDAT$}

Recall that $v : \X \rto \RDA$ is the valuation that we have fixed.
\begin{definition}
  For a susbset $T$ of $\AD$, the the valuation
  $v_{T} : \X \rto \RDAT$ is defined by the formula
  $v_{T}(x) = \interior{v(x)}$, for each $x \in \X$.
\end{definition}
More explicitley, we have
\begin{align*}
  v_{T}(x) & := (\alpha, \closure[\alpha]{T \cap X})\,, \quad
  \text{where } (\alpha,X) = v(x)\,.
\end{align*}

The valuation $v_{T}$ takes values in $\RDAT$, while $v$ takes value
in $\RDA$.  It is possible then to evaluate a lattice term $t$ in
$\RDAT$ using $v_{T}$ and to evaluate it in $\RDA$ using $v$. To
improve readability, we shall use the notation $\evT{t}$ for the
result of evaluating the term in $\RDAT$, 
and the
notation $\evv{t}$ for the result of evaluating it in $\RDA$.
Since both $\evv{t}$ and $\evT{t}$ are subsets of
$P(A \cup X)$, it is possible to compare them using inclusion.

\begin{lemma}
  \label{lemma:monotonicity}
  The relation $\evT{s} \subseteq \evv{s}$ holds, 
  for each $T\subseteq \AD$ and each lattice term $s$.
\end{lemma}
\begin{proof}
  The proof of the Lemma is a straightforward induction, considering
  that $v_{T}(x) \subseteq v(x)$ for all $x \in \X$. 
  For example, using $\evT{s_{i}} \subseteq \evv{s_{i}}$, for
  $i =1,2$,
  \begin{align*}
    \evT{s_{1} \land s_{2}} & = \evT{s_{1}} \mmeet \evT{s_{2}}
    \subseteq
    \evT{s_{1}} \cap \evT{s_{2}}
    \subseteq \evv{s_{1}} \cap \evv{s_{2}} = \evv{s_{1} \land
      s_{2}}\,.
    \tag*{\qedhere}
  \end{align*}
\end{proof}

A straightforward induction also yields:
\begin{lemma}
  \label{lemma:stabilityTwo}
  Let $T\subseteq \AD$ be a finite subset, let $t$ be a lattice term
  and suppose that $\evv{t} = (\beta,Y)$. Then $\evT{t}$ is of the
  form $(\beta,Y')$ for some $Y' \subseteq \AD$.
\end{lemma}
\begin{excludeproof}
  The proof of the Lemma is a straightforward induction. Indeed, the
  Lemma holds if $t$ is a variable. For $t = s_{1} \land s_{2}$ or
  $t = s_{1} \vee s_{2}$ the first component of $\evv{t}$ and
  $\evT{t}$ are computed exactly in the same way, from $\evv{s_{i}}$
  and $\evT{s_{i}}$, respectively.
\end{excludeproof}

\begin{definition}
  \label{lemma:tableau}
  Let us define, for each term $t$ and $f \in \AD$ such that $f \in \evv{t}$, 
  a finite set $T(f,t) \subseteq \AD$ as follows:
\begin{itemize}
\item If $t$ is the variable $x$, then we let $T(f,t) := \set{f}$.
  
\item If $t = s_{1} \land s_{2}$, then
  $f \in \evv{s_{1}}\cap \evv{s_{2}}$, so we define
  $T(f,t) := T(f,s_{1}) \cup T(f,s_{2})$.

\item If $t = s_{1} \vee s_{2}$ and $\evv{s_{i}} = (\alpha_{i},X_{i})$
  for $i = 1,2$, then $f \in \evv{s_{1} \vee s_{2}}$ gives that, for
  some $i \in \set{1,2}$ there exists $g \in X_{i}$ such that
  $\delta(f,g) \subseteq \alpha_{1} \cup \alpha_{2}$. We set then
  $T(f,t) := \set{f} \cup T(g,s_{i})$.
  \end{itemize}
\end{definition}

Obviously, we have:
\begin{lemma}
  \label{lemma:grounded}
  For each lattice term $t$ and $f \in \AD$ such that $f \in \evv{t}$,
  $f \in T(f,t)$.
\end{lemma}
\begin{excludeproof}
  A straightforward induction:
  \begin{itemize}
  \item If $t$ is the variable $x$, so $f \in \evv{x} = v(x)$, then we
    have defined $T(f,x) = $.
  \item If $t = s_{1} \land s_{2}$, then we have defined
    $T(f,t) = T(f,s_{1}) \cup T(f,s_{2})$. From
    $f \in T(f,s_{1}) \cap T(f,s_{2})$ we obtain $f \in T(f,t)$.
  \item If $t = s_{1} \vee s_{2}$, then we have defined
    $T(f,t) = \set{f} \cup T(g,s_{i})$, where $i \in \set{1,2}$ is
    such that $g \in \evv{s_{i}}$ and $\d(f,g) \subseteq
    \evv{s_{1}} \cup \evv{s_{2}}$.
    Clearly, $f \in T(f,t) = \set{f} \cup T(g,s_{i})$.
    \qedhere
  \end{itemize}
\end{excludeproof}
\begin{proposition}
  \label{lemma:tableauTwo}
  For each lattice term $t$ and $f \in \AD$ such that $f \in \evv{t}$,
  if $T(f,t) \subseteq T$, then $f \in \evT{t}$.
\end{proposition}
\begin{proof}
  We prove the statement by induction on $t$.
  \begin{itemize}
  \item If $t$ is the variable $x$ and
    $f \in \evv{x} = v(x) = (\beta,Y)$, then $f \in Y$. We have
    $T(f,x) = \set{f}$. Obviously,
    $f \in Y \cap \set{f} = Y \cap T(f,t) \subseteq Y \cap T$, so
    $f \in (\beta, \closure[\beta]{Y \cap T}) = v_{T}(x) =
    \evT{t}$.
    
  \item Suppose $t = s_{1} \land s_{2}$ so
    $f \in \evv{s_{1} \land s_{2}}$ yields $f \in \evv{s_{1}}$ and
    $f\in\evv{s_{2}}$.
    We have defined $T(f,t) = T(f,s_{1}) \cup T(f,s_{2}) \subseteq T$ and
    so, using $T(f,s_{i}) \subseteq T$ and the induction  hypothesis,
    $f \in \evT{s_{i}}$ for $i = 1,2$. By Lemma~\ref{lemma:grounded}
    $f \in T$, so we can use
    Lemma~\ref{lemma:conj} asserting that
    \begin{align*}
      f \in \evT{s_{1}} \;\mmeet\; \evT{s_{2}} & = \evT{s_{1} \land s_{2}}\,.
    \end{align*}
    
  \item Suppose $t = s_{1} \vee s_{2}$ and
    $f \in \evv{s_{1} \vee s_{2}}$; let also
    $(\beta_{i},Y_{i}) := \evv{s_{i}}$ for $i =1,2$.  We have defined
    $T(f,t) := \set{f} \cup T(g,s_{i})$ for some $i \in \set{1,2}$ and
    for some $g \in \evv{s_{i}}$ such that
    $\d(f,g) \subseteq \beta_{1} \cup \beta_{2}$. Now $g \in
    T(g,s_{i}) \subseteq T(f,t) \subseteq T$ so, by the induction
    hypothesis, 
    $g \in \evT{s_{i}}$.
    According to Lemma~\ref{lemma:stabilityTwo}, for each $i = 1,2$
    $\evT{s_{i}}$ is of the form $(\beta_{i},Y_{i}')$, for some subset
    $Y'_{i} \subseteq \AD$. Therefore
    $\d(f,g) \subseteq \beta_{1} \cup \beta_{2}$ and
    $g \in \evT{s_{i}}$ implies
    \begin{align*}
      f \in \evT{s_{1}} \vee \evT{s_{2}} = \evT{s_{1} \vee s_{2}}\,.
      \tag*{\qedhere}
    \end{align*}
  \end{itemize}
\end{proof}

\begin{proposition}
  \label{prop:inLT}
  Suppose $f$ witnesses the failure of the inclusion $t \leq s$ in
  $\RDA$ w.r.t. the valuation $v$. Then, for each subset
  $T \subseteq \AD$ such $T(f,t) \subseteq T$, $f$ witnesses the
  failure of the inclusion $t \leq s$ in the lattice $\RDAT$ and
  w.r.t. valuation $v_{T}$.
\end{proposition}
\begin{proof}
  As $f$ witnesses $t \not\leq s$ in $\RDA$,
  $f \in \evv{t}$ and $f \not \in \evv{s}$.  By
  Lemma~\ref{lemma:tableauTwo} $f \in \evT{t}$. If $f\in \evT{s}$,
  then $\evT{s} \subseteq \evv{s}$ (Lemma~\ref{lemma:monotonicity})
  implies $f \in \evv[T]{s}$, a contradicition. Therefore
  $f\not\in \evT{s}$, so $f$ witnesses $t \not\leq s$ in $\RDAT$.
  \qed
\end{proof}

\subsection{Preservation of the failure in a finite lattice
  $\L(X,\d)$}

From now on, we suppose that $T \subseteq \AD$ is finite and
$T(f,t) \subseteq T$ with $f$ witnessing the failure of $t \leq s$.
Consider the sub-Boolean-algebra of $P(A)$ generated by the sets
\begin{align}
  \label{eq:generators}
  & \set{ \d(f,g) \mid f,g \in T } \cup \set{A \cap v(x) \mid
  x \in Vars(t,s)}\,.
\end{align}
Let us call $\BT$ this Boolean algebra (yet, notice the dependency of
this definition on $T$, as well as on $t,s$ and $v$). It is well known
that a Boolean algebra generated by a finite set is finite.
\begin{remark}
  \label{rem:sizeAtB}
  If $n = \card(T)$ and $m = \card(Vars(t,s))$, then $\BT$ can have at
  most $2^{\frac{n(n-1)}{2} + m}$ atoms.  If we let $k$ be the maximum
  of the sizes of $t$ and $s$, then, for $T = T(f,t)$, both $n \leq k$
  and $m \leq 2k$. We obtain in this case the over-approximation
  $2^{\frac{k^{2} + 3k}{2}}$ on the number of atoms of $\B$.
\end{remark}
Let us also recall that $\BT$ is isomorphic to the powerset $P(\AB)$,
where $\AB$ is the set of atoms of $\BT$. Let $i : P(\AB) \rto P(A)$
be an injectve homomorphism of Boolean algebras whose image is $\B$.
Since $\d(f,g) \in \BT$ for every $f,g \in T$, we can transform the
metric space $(T,\d)$ induced from $(\AD,\d)$ into a metric space
$(T,\d_{\AB})$ whose distance takes values in the powerset algebra
$P(\AB)$:
\begin{align*}
  \d_{\AB}(f,g) = \beta & \quad \text{if and only if} \quad \d(f,g) = i(\beta)\,.
\end{align*}
Recall from Proposition~\ref{prop:changeOfBA} that there is a lattice
embedding $\is : \L(T,\d_{\AB}) \rto \L(T,\d)$, defined in the obvious
way: $\is(\alpha,Y) = (i(\beta),Y)$.

\begin{proposition}
  \label{prop:finitization}
  If $f$ witnesses the failure of the inclusion $t \leq s $ in $\RDA$
  w.r.t. the valuation $v$, then the same inclusion fails in all the
  lattices $\L(T,\dAB)$, where $T$ is a finite set and
  $T(f,t) \subseteq T$.
\end{proposition}
\begin{proof}
  By Proposition~\ref{prop:inLT} the inclusion $t \leq s$ fails in the
  lattice $\RDAT$. This lattice is isomorphic to the lattice
  $\L(T,\d)$ via the map sending $(\alpha,X) \in \RDAT$ to
  $(\alpha,X \cap T)$. Up to this isomorphism, it is seen that the
  (restriction to the variables in $t$ and $s$ of) the valuation
  $v_{T}$ takes values in the image of the lattice $\L(T,\dAB)$ via
  $\is$, so $\evT{t},\evT{s}$ belong to this sublattice and the
  inclusion fails in this lattice, and therefore also in $\L(T,\dAB)$.
  \qed
\end{proof}

\section{Preservation of the failure in a finite lattice $\L(\Secpi)$}
\label{sec:completion}

We have seen up to now that if $t \leq s$ fails in $\RDA$, then it
fails in many lattices of the form $\L(T,\dAB)$. Yet it is not obvious
a priori that any of these lattices belongs to the variety generated
by the relational lattices. We show in this section that we can extend
any $T$ to a finite set $\GT$ while keeping $\B$ fixed, so that
$(\GT,\dAB)$ is a \Pc space over $P(\AB)$. Thus, the inclusion
$t \leq s$ fails in the finite lattice $\L(\GT,\dAB)$. Since
$(\GT,\dAB)$ is isomorphic to a space of the form $\Secpi$ with
$\pi : E \rto \AB$, the inclusion $t \leq s$ fails in a lattice
$\L(\Secpi)$ which we have seen belongs to the variety generated by
the relational lattices.  This also leads to construct a finite
relational lattice $\R(\AB,E)$ in which the equation $t \leq s$ fails.
By following the chain of constructions, the sizes of $\AB$ and $E$
can also be estimated, leading to decidability of the equational
theory of relational lattices.

\begin{definition}
  \label{def:glues}
  A \emph{glue of $T$ and $\B$} is a function $g \in \AD$ such that,
  for all $\alpha \in \AB$, there exists $f \in T$ with
  $f \restr \alpha = g$.  We denote by $\GT$ the set of all functions
  that are glues of $T$ and $\B$.
\end{definition}
Observe that $T \subseteq \GT$ and that $\GT$ is finite,
with 
\begin{align}
  \label{eq:cardGlue}
  \card(\GT) & \leq \card(T)^{\card(\AB)} \,.
\end{align}

In order to prove the following Lemma, let, for each $\alpha \in \AB$
and $g \in \GT$, $f(g,\alpha) \in T$ be such that
$g \restr \alpha = f(g,\alpha) \restr\alpha$.
\begin{lemma}
  If $g_{1},g_{2} \in \GT$, then $\d(g_{1},g_{2}) \in \B$.
\end{lemma}
\begin{proof}
  \begin{align*}
    \d(g_{1},g_{2}) & = \bigcup_{\alpha \in \AB} (\alpha \cap
    \d(g_{1},g_{2}))
    = \bigcup_{\alpha \in \AB} (\alpha \cap
    \d(f(g_{1},\alpha),f(g_{2},\alpha)))\,.
  \end{align*}
  Since $\d(f(g_{1},\alpha),f(g_{2},\alpha)) \in \B$ and $\alpha$ is
  an atom of $\B$, each expression of the form
  $\alpha \cap \d(f(g_{1},\alpha),f(g_{2},\alpha))$ is either
  $\emptyset$ or $\alpha$. It follows that $\d(g_{1},g_{2}) \in \B$.
  \qed
\end{proof}

For a Boolean subalgebra $B$ of $P(A)$, we say that a subset $T$ of
$\AD$ is \emph{\Pc relative to $B$} if, for each $f,g \in T$,
\begin{enumerate}
\item $\d(f,g) \in B$,
\item $\d(f,g) \subseteq \beta\cup \gamma$, implies
  $\d(f,h) \subseteq \beta$ and $\d(h,g) \subseteq \gamma$ for some
  $h \in T$, for each $\beta,\gamma \in B$.
\end{enumerate}

\begin{lemma}
  The set $\GT$ is \Pc relative to the Boolean algebra $\B$.
\end{lemma}
\begin{proof}
  Let $f,g \in \GT$ be such that
  $\d(f,g) \subseteq \beta \cup \gamma$.  Let $h \in \AD$ be defined
  so that, for each $\alpha \in \AB$,
  $h \restr \alpha = f \restr \alpha$ if $\alpha \not\subseteq \beta$
  and $h \restr \alpha = g \restr \alpha$, otherwise.  Obviously,
  $h \in \GT$.

  Observe that $\alpha \not\subseteq \beta$ if and only if
  $\alpha \subseteq \beta^{\compl}$, for each $\alpha \in \AB$, since
  $\beta \in \B$.  We deduce therefore
  $h \restr \alpha = f \restr \alpha$ if $\alpha \in \AB$ and
  $\alpha \subseteq \beta^{\compl}$, so $f(a) = h(a)$ for each
  $a \in \beta^{\compl}$.  Consequently
  $\beta^{\compl} \subseteq Eq(f,h)$ and $\d(f,h) \subseteq \beta$.

  We also have $h \restr \alpha = g \restr \alpha$ if $\alpha \in \AB$
  and $\alpha \subseteq \gamma^{\compl}$. As before, this implies
  $\d(h,g) \subseteq \gamma$.  Indeed, this is the case if
  $\alpha \subseteq \beta$, by definition of $h$. Suppose now that
  $\alpha \not\subseteq \beta$, so
  $\alpha \subseteq \beta^{\compl} \cap \gamma^{\compl} = (\beta \cup
  \gamma)^{\compl}$. Since $\d(f,g) \subseteq \beta \cup \gamma$, then
  $\alpha \subseteq \d(f,g)^{\compl} = Eq(f,g)$, i.e.
  $f \restr \alpha = g \restr \alpha$. Together with
  $h \restr \alpha = f \restr \alpha$ (by definition of $h$) we obtain
  $h \restr \alpha = f \restr \alpha$.
  \qed
\end{proof}

We can finally bring together the observations developed so far and
state our main results.
\begin{theorem}
  \label{theo:main1}
  If an inclusion $t \leq s $ fails in all the lattices $\RDA$, then
  it fails in a finite lattice $\R(E,A')$, where
  $\card(A') \leq 2^{p(k)}$ with $k = \max(size(t),size(s))$,
  $p(k) = \frac{2^{k^{2}} + 3k}{2}$, and $\card(E) \leq size(t)$.
\end{theorem}
\begin{proof}
  By Proposition~\ref{prop:inLT} the inclusion $t \leq s$ fails in all
  the lattices $\RDAT$ where $T(f,t) \subseteq T$. Once defined $\B$
  as the Boolean subalgebra of $P(A)$ generated by the sets as in the
  display \eqref{eq:generators} (with $T = T(f,T)$) and $\GT$ as the
  set of glues of $T(f,t)$ and $\B$ as in Definition~\ref{def:glues},
  the inclusion fails in $\RDA_{\GT}$, since $T(f,T) \subseteq \GT$,
  and then in $\L(\GT,\dAB)$ by
  Proposition~\ref{prop:finitization}. The condition that $\GT$ is \Pc
  relative to $B$ is equivalent to saying that the space $(\GT,\dAB)$
  is \Pc. This space is therefore isomorphic to a space of the form
  $\Secpi$ for some surjective $\pi : F \rto \AB$, and $t \leq s$
  fails in $\L(\Secpi)$.

  Equation~\eqref{eq:cardGlue} shows that, for each $\alpha \in \AB$,
  $F_{\alpha} = \pi^{-1}(\alpha)$ has cardinality at most
  $\card(T(f,t))$ and the size of $t$ is an upper bound for
  $\card(T(f,t))$. We can therefore embed the space $\Secpi$ into a
  space of the form $(E^{\AB},\d)$ with the size of $t$ an upper bound
  for $\card(E)$. The proof of Proposition~\ref{prop:HSSecpi} exhibits
  $\L(\Secpi)$ as a homomorphic image of a sublattice of
  $\L(E^{\AB},\d)$ and therefore the inclusion $t \leq s$ also fails
  within $\L(E^{\AB},\d) = \R(E,\AB)$.  The upper bound on the size of
  $\AB$ has been extimated in Remark~\ref{rem:sizeAtB}. \qed
\end{proof}

\begin{remark}
  In the statement of the previous Theorem, the size of the lattice
  $\R(E,A')$ can be estimated out of the sizes of $E$ and $A'$
  considering that
  $$
  P(\expo{A'}{E}) \subseteq \R(E,A') \subseteq P(A' \cup
  \expo{A'}{E})\,.
  $$
  An upper bound for $\card(\R(E,A'))$ is therefore
  $2^{p(k) + k^{2^{p(k)}}}$ where $p(k)$ is the polynomial of degree
  $2$ as in the statement of the Theorem and $k$ is the maximum of
  $size(t),size(s)$.
\end{remark}

A standard argument yields now:
\begin{corollary}
  The equational theory of the relational lattices is decidable.
\end{corollary}

\section{Conclusions}
\label{sec:conclusions}
We argued that the equational theory of relational lattices is
decidable. We achieved this goal by giving a finite (counter)model
construction of bounded size.

Our result leaves open other questions that we might ask on relational
lattices. We mentioned in the introduction the quest for a complete
axiomatic base for this theory or, anyway, the need of a complete
deductive system---so to develop automatic reasoning for the algebra
of relational lattices. We are confident that the mathematical
insights contained in the decidability proof will contribute to
achieve this goal.

Our result also opens new research directions, in primis, the
investigation of the complexity of deciding lattice-theoretic
equations/inclusions on relational lattices. Of course, the obvious
decision procedure arising from the finite model construction is not
optimal; few algebraic considerations already suggest how the decision
procedure can be improved.

Also, it would be desirable next to investigate decidability of
equational theories in signatures extending of the pure lattice
signature; many such extensions are proposed in \cite{LitakMHjlamp}.
It is not difficult to adapt the present decidability proof so to add
to the signature the header constant.

A further interesting question is how this result translates back to
the field of multidimensional modal logic \cite{Kurucz2007}. We
pointed out in \cite{San2016} how the algebra of relational lattices
can be encoded into multimodal framework; we conjecture that our
decidability result yields the decidability of some positive fragments
of well known undecidable logics, such as the products $\Sfive^{n}$
with $n \geq 3$.

\ifdefined\llncs
\newpage
\else\fi

\bibliographystyle{abbrv}
\bibliography{biblio}

\end{document}